\documentstyle[aps]{revtex}

\begin{document}
\author{S. Shelly Sharma$^{1}$, N. K. Sharma$^{2}$}
\address{$^{1}$Depto. de Fisica, Universidade Estadual de Londrina, Londrina,\\
86040-370,\\
PR, Brazil}
\address{$^{2}$Depto. de Matematica, Universidade Estadual de Londrina, Londrina,\\
86040-370, PR, Brazil}
\title{Schr\"{o}dinger cat state of trapped ions in harmonic and anharmonic
oscillator traps\\
}
\maketitle

\begin{abstract}
We examine the time evolution of a two level ion interacting with a light
field in harmonic oscillator trap and in a trap with anharmonicities. The
anharmonicities of the trap are quantified in terms of the deformation
parameter $\tau $ characterizing the $q$-analog of the harmonic oscillator
trap. Initially the ion is prepared in a Schr\"{o}dinger cat state. The
entanglement of the center of mass motional states and the internal degrees
of freedom of the ion results in characteristic collapse and revival
pattern. We calculate numerically the population inversion $I(t)$,
quasi-probabilities $Q(t),$ and partial mutual quantum entropy $S(P),$ for
the system as a function of time. Interestingly, small deformations of the
trap enhance the contrast between population inversion collapse and revival
peaks as compared to the zero deformation case. For $\beta =3$ and $4,$( $%
\beta $ determines the average number of trap quanta linked to center of
mass motion) the best collapse and revival sequence is obtained for $\tau
=0.0047$ and $\tau =0.004$ respectively. For large values of $\tau $
decoherence sets in accompanied by loss of amplitude of population inversion
and for $\tau $ $\sim 0.1$ the collapse and revival phenomenon disappear.
Each collapse or revival of population inversion is characterized by a peak
in $S(P)$ versus $t$ plot. During the transition from collapse to revival
and vice-versa we have minimum mutual entropy value that is $S(P)=0.$
Successive revival peaks show a lowering of the local maximum point
indicating a dissipative irreversible change in the ionic state. Improved
definition of collapse and revival pattern as the anharminicity of the
trapping potential increases is also reflected in the Quasi- probability
versus $t$ plots.
\end{abstract}


\section{ \ Introduction}

Two-level ions trapped in a harmonic or anharmonic trap have turned into an
important tool for understanding the time evolution of non classical states.
Experimentally harmonic oscillator traps have been realized and various
types of non-classical states of ions in trapped systems constructed \cite
{Meek96,Monr296} with sufficient control over amplitudes, relative phases
etc. of the component states. It is now possible to detect\cite{Lieb96}
experimentally constructed non-classical states and measure the statistics
of the quantum motion of the center of mass of the ion. The non-classical
states constructed and detected include the Fock states, thermal states,
Shrodinger cat states and squeezed states. The possibility of manipulating
the entangled states of particle systems provides the basis for applications
like quantum computations and quantum communications. On the other hand,
these experimental advances have paved a way to an understanding of some of
the basic aspects of quantum theory. A controlled manipulation of these
states requires precise control of the Hamiltonian of the system. The form
of the trap potential is usually controlled by the geometry of the system,
the laser field intensities and relative phases. As such anharmonic traps
with different potential shapes should also be viable experimentally. The
measurement of dynamical behavior of non-classical states in harmonic and
anharmonic traps and a study of anharmonicity related new features makes an
interesting study. Harmonically trapped and laser driven ion system has been
studied in many contexts such as nonlinear coherent states\cite
{Mato96pra,Shiv99}, vibrionic Jaynes-Cummings interactions\cite{Bloc92},
nonlinear Jaynes-Cummings interactions\cite{Voge89,Voge95}, and generation
of even and odd coherent states\cite{Mato96}. The trapped ion-laser system
is in some sense an experimental realization of Jaynes Cummings model \cite
{Jayn63}, just like the micromaser cavity experiments. A study of the
dynamics of two-level atom coupled to the $q$-analog of a single mode of the
bosonic cavity field \cite{Buze92} indicates that $q$ deformation of
Heisenberg algebra may correspond to some effective nonlinear interaction of
the cavity mode. In our earlier work \cite{She97} on ions in harmonic and
anharmonic traps, we examined the collapses and revivals of the population
inversion and quasi-probabilities as a function of time for the initial
state of the system having ion in it's ground state and the center of mass
motion described by a coherent state. This study brought out some very
interesting new features in the dynamics of the system. In the present work,
we consider a two-level ion initially in a Schr\"{o}dinger cat state and
examine the collapse and revival of population inversion and
quasi-probabilities with a focus on the coherence of the system and
anharmonicities of the trap. We also examine the time evolution of partial
mutual entropy of the system to understand the improvement in collapse and
revival pattern of the system for characteristic values of parameter
quantifying the trap anharmonicities. The motivation for studying these
states is provided by the possible use in quantum computation and quantum
communication where entanglement of ionic internal states with center of
mass motion may be used to encode information \cite
{Benn92,Benn93,Eker91,Eker92}. In Ref. \cite{She97} as well as in this work,
the anharmonicity of the trap potential is modeled through a $q$-analog
harmonic oscillator trap. Since Macfarlane \cite{Macf89} and Biedenharn \cite
{Bied89} discussed $q$-analog of harmonic oscillator, a lot of work using
these in various fields of physics has been done. In our efforts to
understand the physical nature of $q$-deformations in the context of pairing
effect in nuclei\cite{She92,She94,She00}, the $q$-deformation is found to
simulate the nonlinearities of the problem caused by unaccounted part of the
residual interaction. In the present context, with the parameter $q$ being
used to quantify the anharmonicity of the trap potential, the underlying
energy spectrum of vibrational states of ion as well as the laser ion
interaction contain dependence on the number operator for vibrational quanta
and the parameter $q$. For small values of $q$, the $q$-analog harmonic
oscillator is essentially an anharmonic oscillator with $x^{6}$
anharmonicities \cite{Bona99}. Nonlinear couplings of vibrational modes of
ion have been studied theoretically \cite{Wall9798,Agar97,Stein97} with the
nonlinear effects modeled in the laser-ion interaction Hamiltonian through
Hermitian operators with strong dependence on the number operator for
vibrational quanta. The underlying trap in these studies is a harmonic
oscillator one. We may point out that for ion in a harmonic oscillator trap
and non-linear ion-laser coupling expressed through $q-$deformed operators
with the $q$ values used in our work, the system behavior continues to be
very close to the case of ion in a harmonic trap and linear ion-laser
couplings.

We start with a description of the trapped ion-laser system Hamiltonian in
section I for the cases of ion in a Harmonic Oscillator trap and ion in a $q$%
-analog harmonic oscillator trap. In section II the initial state and the
equations governing it's time evolution are given. Population inversion and
quasi-probability are defined in section III. In section IV, we discuss the
concept of partial mutual entropy for the system and the numerical results
are analyzed in section V.

\section{The Hamiltonian}

\subsection{Harmonic Oscillator trap}

The system that consists of the two level ion moving in a harmonic
oscillator trap potential and interacting with a classical single-mode light
field of frequency $\omega _{l}$ is described by the Hamiltonian, 
\begin{equation}
H=\frac{1}{2}\hbar \omega (a^{\dagger }a+aa^{\dagger })+\frac{1}{2}\hbar
\triangle \sigma _{z}+\frac{1}{2}\hbar \Omega (F\sigma ^{+}+F^{*}\sigma ^{-})
\label{eq1}
\end{equation}
where $a$ and $a^{\dagger }$ are the usual creation and destruction
operators for the oscillator, $\Delta =\omega _{a}-\omega _{l}$ , is the
detuning parameter and $\Omega $ is the Rabi frequency of the system. The
pseudo-spin operators of the two-level ion, $\sigma ^{\pm }$ and $\sigma _{z}
$ operate on state vectors $\left| g\right\rangle $ and $\left|
e\right\rangle $ where indices $g$ and $e$ stand for the ground and excited
states of the two level ion. The operator $F$ stands for\cite{Bloc92} $\exp
(ikx)=\exp [i\epsilon (a^{\dagger }+a)]$, where $k$ is the wave vector of
the light field and $x$ position quadrature of the center of mass. The
parameter $\epsilon =\sqrt{\frac{E_{r}}{E_{t}}}$ is a function of the ratio
of the recoil energy of the ion $E_{r}=\frac{\hbar ^{2}k^{2}}{2m}$and the
characteristic trap quantum energy $E_{t}=\hbar \omega =\frac{v}{a_{0}}$.
The second term in the Hamiltonian refers to the energy associated with
internal degrees of freedom of the ion, whereas the third term is the
interaction of the ion with the light field. By using a special form of
Baker-Hausdorff Theorem the operator $F$ may be written as a product of
operators i.e.e we use the equality

\begin{equation}
F=\exp [i\epsilon (a^{\dagger }+a)]=e^{(\frac{\left| \epsilon \right| ^{2}}{2%
}[a^{\dagger },a])}e^{i\epsilon a^{\dagger }}e^{i\epsilon a}.  \label{eq2}
\end{equation}
For harmonic oscillator trap one has 
\begin{equation}
\lbrack a,a^{\dagger }]=1\medskip \ ;\medskip \ \ \ Na^{\dagger }-a^{\dagger
}N=a^{\dagger }\medskip \ ;\medskip \ Na-aN=-a.  \label{eq3}
\end{equation}

The physical processes implied by the various terms of the operator 
\begin{equation}
F=\left[ \exp \left( \frac{-\epsilon ^{2}}{2}\right) \sum_{n}\frac{\left(
i\epsilon \right) ^{n}a^{\dagger n}}{n!}\sum_{k}\frac{\left( i\epsilon
\right) ^{k}a^{k}}{k!}\right] ,  \label{eq4}
\end{equation}
may be divided into three categories. The terms for $n>k$ correspond to an
increase in energy linked with the motional state of center of mass of the
ion by $(n-k)$ quanta. The terms with $n<k$ represent destruction of $(k-n)$
quanta of energy thus reducing the amount of energy linked with the center
of mass motion. For $(n=k),$ we have diagonal contributions. The
contribution from a particular term containing operators $a^{\dagger n}a^{k}$%
is determined by the coefficient $\exp \left( \frac{-\epsilon ^{2}}{2}%
\right) \left[ \frac{\left( i\epsilon \right) ^{n+k}}{n!k!}\right] $ .

\subsection{$q$-analog harmonic oscillator trap}

Next we consider a single two level ion having ionic transition frequency$\
\omega _{a}$ in a quantized $q$-analog quantum harmonic oscillator trap ($q$%
-deformed harmonic oscillator trap) interacting with a single mode
travelling light field. The creation and annihilation operators for the trap
quanta satisfy the following commutation relations,

\begin{equation}
AA^{\dagger }-qA^{\dagger }A=q^{-N}\hspace{0.2in};\hspace{0.2in}NA^{\dagger
}-A^{\dagger }N=A^{\dagger }\hspace{0.2in};\hspace{0.2in}NA-AN=-A.
\label{eq.5}
\end{equation}
The operators $A$ and $A^{\dagger }$ act in a Hilbert space with basis
vectors $\left| n\right\rangle $, $n=0,1,2,...,$ given by,

\begin{equation}
\left| n\right\rangle =\frac{(A^{\dagger })^{n}}{([n]_{q}!)^{\frac{1}{2}}}%
\left| 0\right\rangle  \label{eq6}
\end{equation}
such that the number operator $N\left| n\right\rangle =n\left|
n\right\rangle .$ The vacuum state is $A\left| 0\right\rangle =0.$ We define
here $[x]_{q}$ as 
\begin{equation}
\lbrack x]_{q}=\frac{q^{x}-q^{-x}}{q-q^{-1}}  \label{eq7}
\end{equation}
and the $q$-analog factorial $[n]_{q}!$ is recursively defined by

[0]$_{q}!=[1]_{q}!=1$ and $[n]_{q}!=[n]_{q}[n-1]_{q}!.$ It is easily
verified that 
\begin{equation}
A^{\dagger }\left| n\right\rangle =[n+1]_{q}^{\frac{1}{2}}\left|
n+1\right\rangle \medskip \ ;\medskip \ \qquad A\left| n\right\rangle
=[n]_{q}^{\frac{1}{2}}\left| n-1\right\rangle  \label{eq8}
\end{equation}
and $N$ is not equal to $A^{\dagger }A.$ Analogous to the harmonic
oscillator one may define the $q$-momentum and $q$-position coordinate 
\begin{equation}
P_{q}=i\sqrt{\frac{m\hbar \omega }{2}}(A^{\dagger }-A)\medskip \ ;\medskip
\qquad \ X_{q}=\sqrt{\frac{\hbar }{2m\omega }}(A^{\dagger }+A).  \label{eq9}
\end{equation}
The $q$-analog harmonic oscillator Hamiltonian is given by 
\begin{equation}
H_{qho}=\frac{1}{2}\hbar \omega (AA^{\dagger }+A^{\dagger }A)  \label{eq10}
\end{equation}
with eigenvalues 
\begin{equation}
E_{n}=\frac{1}{2}\hbar \omega ([n+1]_{q}+[n]_{q}).  \label{eq11}
\end{equation}
We note that the trap states are not evenly spaced, the energy spacing being
a function of deformation. Besides that as we move up in the number of
vibrational quanta in the states the spacing between successive states
increases. We choose $q=e^{\tau },$ where $\tau $ is a real valued
parameter. For $\tau =0.0$ the harmonic oscillator trap is recovered. As
such the parameter $\tau $ is a measure of the extent to which the harmonic
oscillator potential trap is deformed. Using the Taylor expansion of the $q$
- number $[n]_{q}$ in terms of powers of $\tau ^{2}$ the energy eigenvalues
may be rewritten for small $\tau $ as 
\begin{equation}
E_{n}=\hbar \omega \left[ \left( n+\frac{1}{2}\right) \left( 1-\frac{\tau
^{2}}{24}\right) +\left( n+\frac{1}{2}\right) ^{3}\frac{\tau ^{2}}{6}%
+....\right] .  \label{eq12}
\end{equation}
Bonatsos et. Al \cite{Bona94} have shown that the potential giving a
spectrum similar to that of Eq. (\ref{eq12}) up to the order $\tau ^{2}$
looks like 
\begin{equation}
V(x)=\left( \frac{1}{2}-\frac{\tau ^{2}}{8}\right) x^{2}+\frac{\tau ^{2}}{120%
}x^{6},  \label{eq12a}
\end{equation}
which is an anharmonic oscillator with $x^{6}$ anharmonicities \cite{Bona99}.

Using a nonlinear map given by Curtright and Zachos \cite{Curt90} one can
express $H_{qho\text{ }}$in terms of the operators $a$ and $a^{\dagger }$ of
the harmonic oscillator. To make this point clear we express the operators $%
A $ and $A^{\dagger }$ as

\begin{equation}
A=a\,f(N)\qquad ;\qquad A^{\dagger }=f(N)\,a^{\dagger }  \label{eq13}
\end{equation}

where $f(N)=\left( \frac{[N]_q}N\right) ^{\frac 12}$ and $N=a^{\dagger }a$.
We can also verify that

\begin{equation}
f(N)a^{\dagger }=a^{\dagger }f(N+1)\medskip \ ;\medskip \ f(N)a=af(N-1)
\label{eq14}
\end{equation}
The Hamiltonian of Eq. \ref{eq10} can be rewritten as 
\begin{equation}
H_{qho}=\hbar \omega \,\left( f(N+1)^{2}+f(N)^{2}\right) \left( a^{\dagger
}a+\frac{1}{2}\right)  \label{eq15}
\end{equation}
which can be interpreted as a harmonic oscillator Hamiltonian with a
frequency $\omega _{q}(N)$ that depends on the quantum number $n$ of the
state in question.

The quadratures $x$ and $p$ are related to $P_{q}$ and $\ X_{q}$ through 
\begin{eqnarray}
P_{q} &=&i\sqrt{\frac{m\hbar \omega }{2}}\left( \left( f(N)-f(N+1)\right) 
\frac{x}{\sqrt{2}}-\left( f(N)+f(N+1)\right) \frac{ip}{\sqrt{2}}\right) 
\nonumber \\
X_{q} &=&\sqrt{\frac{\hbar }{2m\omega }}\left( \left( f(N)+f(N+1)\right) 
\frac{x}{\sqrt{2}}-\left( f(N)-f(N+1)\right) \frac{ip}{\sqrt{2}}\right)
\label{eq15b}
\end{eqnarray}

\subsection{Hamiltonian for an ion interacting with light in a $q$-analog
harmonic oscillator trap}

The Hamiltonian for an ion interacting with light in a $q$-analog harmonic
oscillator trap may now be written as

\[
H_{q}=\frac{1}{2}\hbar \omega (AA^{\dagger }+A^{\dagger }A)+\frac{1}{2}\hbar
\triangle \sigma _{z}+\frac{1}{2}\hbar \Omega (F_{q}\sigma
^{+}+F_{q}^{\dagger }\sigma ^{-}) 
\]
where by analogy with Eq. \ref{eq4} we choose

\begin{equation}
F_{q}=e^{\left( \frac{-\left| \epsilon \right| ^{2}}{2}\right) }e^{i\epsilon
A^{\dagger }}e^{i\epsilon A}\text{ ,}  \label{eq16}
\end{equation}
which in the limit $q\rightarrow 1$ reduces to $F$, and can be expanded as 
\cite{She99}, 
\begin{equation}
F_{q}=e^{\left( -\frac{\left| \epsilon \right| ^{2}}{2}\right)
}\sum_{n=0}^{\infty }\frac{(i\epsilon )^{n}A^{\dagger }{}^{n}}{n!}%
\sum_{k=0}^{\infty }\frac{(i\epsilon )^{k}A^{k}}{k!}.  \label{eq17}
\end{equation}
Various terms in the expansion of this operator represent processes which
might result in transitions of the center of mass from a given motional
state, in the $q$-analog trap, to another, while loosing or gaining energy.
The coefficient of the operators $A^{\dagger }{}^{n}A^{k}$ is again $\exp
\left( \frac{-\epsilon ^{2}}{2}\right) \left[ \frac{\left( i\epsilon \right)
^{n+k}}{n!k!}\right] $, the same as that in the corresponding term with
operators $a^{\dagger n}a^{k}$ in Eq. \ref{eq4}. Intuitively this is the
correct way of representing the interaction of the ion and the laser in a $q$%
-analog harmonic oscillator trap. This form of the operator shows that the
energy exchange of the center of mass motion of the ion occurs as the ion
moves up or down in the trap. Since $E_{n+1}-E_{n}$ is $n$ dependent, it
implies an $n$ dependent entanglement of the center of mass motion and the
internal degrees of freedom of the two level atom. Using Eq. \ref{eq13}, we
can rewrite Eq. \ref{eq17} as 
\begin{equation}
F_{q}=e^{\left( -\frac{\left| \epsilon \right| ^{2}}{2}\right)
}\sum_{n=0}^{\infty }\sum_{k=0}^{\infty }\frac{(i)^{n+k}\left( \epsilon
f(N)a^{\dagger }\right) {}^{n}\left( \epsilon af(N)\right) ^{k}}{n!k!}.
\label{eq18}
\end{equation}

Comparing with $F$ we notice that the effective lamb Dicke parameter in a $q$%
-analog trap, for the loss and gain of motional state energy in an
interaction process, is $\epsilon f(N)$ where $N$ is the number operator.
With the $q$-analog of Glauber coherent state defined as 
\[
\left| \alpha \right\rangle _{q}=\frac{1}{\sqrt{\exp _{q}^{\left| \alpha
\right| ^{2}}}}\sum\limits_{n=0}^{\infty }\frac{\alpha ^{n}}{\sqrt{\left[
n\right] _{q}!}}\left| n\right\rangle \text{ ,} 
\]
the calculated value of $_{q}\left\langle \alpha \right| f(N)\left| \alpha
\right\rangle _{q}$ is $1.0004$, for the choice $\alpha =4$ , $\epsilon
=0.05 $ and $\tau =0.003$.

It is important to recall at this point that the operator $(A^{\dagger }+A)$
is not a self adjoint operator. The vectors

$\left| m\right\rangle $ are analytic vectors of the operators $(A^{\dagger
})^{n}A^{k}$ but not of the operators $A^{k}(A^{\dagger })^{n}$. One can
easily verify that the matrix elements $\left\langle m\right| e^{\left( 
\frac{\left| \epsilon \right| ^{2}}{2}\right) }\sum_{l=0}^{\infty }\frac{%
(i\epsilon )^{k}A^{k}}{k!}\sum_{n=0}^{\infty }\frac{(i\epsilon
)^{n}A^{\dagger }{}^{n}}{n!}\left| m\right\rangle $ are not well defined. On
the other hand the matrix elements of the operator in Eq. \ref{eq16} are
well defined as discussed in Ref. \cite{She99}. From the form of the
operator in Eq. \ref{eq17} it is clear that the product $F_{q}\sigma ^{+}$
contains all the processes in which the atom is excited from the ground
state to the excited state while at the same time some of the energy quanta
in the center of mass motion are lost or gained. It is also important to
note that $X_{q}$ is not the true position coordinate but is related to $x$
and $p$ in some complex way. As such the operator $\exp [i\epsilon
(A^{\dagger }+A)]$ can not be used to represent the laser-ion interaction.

The expression for the matrix element of the operator $F_{q}$ between the
states $\left\langle m\right| $ and $\left| n\right\rangle $ for $m\leq n$
is given by,

\begin{equation}
\left\langle m\right| F\left| n\right\rangle =\frac{e^{-\frac{\left|
\epsilon \right| ^{2}}{2}}(i\epsilon )^{n-m}[m]_{q}^{\frac{1}{2}}!}{[n]_{q}^{%
\frac{1}{2}}!}\sum\limits_{k=0}^{m}\frac{(\epsilon )^{2k}(-1)^{k}[n]_{q}!}{%
k!(n-m+k)![m-k]_{q}!}\text{ .}  \label{eq19}
\end{equation}

\section{Schr\"{o}dinger Cat State}

Initially the ion is prepared in a Schr\"{o}dinger Cat State with equal
probabilities of finding the ion in its ground state and in the excited
state coupled to a coherent state describing the state of motion of the
center of mass of the ion in the anharmonic trap. The experimental technique
for producing trapped ion in a state with these characteristics has been
described in many experimental papers and such states have been produced 
\cite{Monr296,Monr95} .

The initial state of the trapped ion in a $q$-analog harmonic oscillator
trap can be expressed as

\begin{equation}
\Psi (t=0)=\frac{\left| g,\beta \right\rangle _{q}+e^{i\phi }\ \left|
e,-\beta \right\rangle _{q}}{\sqrt{2}}  \label{eq20}
\end{equation}
where $\left| g,\beta \right\rangle _{q}$ is given by 
\begin{equation}
\left| g,\beta \right\rangle _{q}=\frac{1}{\sqrt{\exp _{q}^{\left| \beta
\right| ^{2}}}}\sum\limits_{n=0}^{\infty }\frac{\beta ^{n}}{\sqrt{\left[
n\right] _{q}!}}\left| g,n\right\rangle ,  \label{eq21}
\end{equation}
and $\beta $ determines the average number of vibrational quanta linked to
the center of mass motion. The relative phase between two possible internal
states is determined by $\phi $, in the present study we choose $e^{i\phi
}=1.$

The state of the system at a time $t$, 
\begin{equation}
\Psi (t)=\sum\limits_{m}g_{m}(t)\left| g,m\right\rangle
+\sum\limits_{m}e_{m}(t)\left| e,m\right\rangle   \label{eq22}
\end{equation}
is a solution of the time dependent Schr\"{o}dinger equation 
\begin{equation}
H_{q}\Psi (t)=i\hbar \frac{d}{dt}\Psi (t).  \label{eq23}
\end{equation}
In the state $\left| g,m\right\rangle $ ($\left| e,m\right\rangle $ ) the
two level ion in it's ground (excited) state is coupled to number state $%
\left| m\right\rangle $ of the anharmonic oscillator. The probability
amplitudes $g_{m}(t)$ and $e_{m}(t)$ satisfy the following set of coupled
equations

\begin{eqnarray}
i\frac{dg_{m}(t)}{dt} &=&\frac{1}{2}g_{m}(t)\left( \omega
([m+1]_{q}+[m]_{q})-\Delta \right) +\frac{1}{2}\Omega
\sum\limits_{n}e_{n}(t)\left\langle g,m\right| \sigma ^{-}F_{q}^{\dagger
}\left| e,n\right\rangle  \nonumber \\
i\frac{de_{m}(t)}{dt} &=&\frac{1}{2}e_{m}(t)\left( \omega
([m+1]_{q}+[m]_{q})+\Delta \right) +\frac{1}{2}\Omega
\sum\limits_{n}g_{n}(t)\left\langle e,m\right| F_{q}\sigma ^{+}\left|
g,n\right\rangle  \label{eq24}
\end{eqnarray}

In the limit $q\rightarrow 1$ we recover the system that gives the dynamics
of two-level ion in a harmonic oscillator trap. By rescaling $\epsilon $, $%
\omega $, and $t$, the parameter $\Omega $ can be eliminated from Eq. (\ref
{eq24}).

\section{Population inversion and Quasi-Probability}

The entanglement of motional degrees of freedom of the center of mass of the
ion with it's internal degrees of freedom manifests itself in the well known
collapse and revival of population inversion. The population inversion is
defined as 
\begin{equation}
I(t)=P_{g}(t)-P_{e}(t),  \label{eq25}
\end{equation}
that is the difference between the probability of finding the system in the
ground state, $P_{g}(t)$, and the probability of finding the system in the
excited state, $P_{e}(t)$. The time evolution of population inversion is
examined for different values of trap anharmonicities for the cases when
coherent state is characterized by parameter $\beta =3,4$ . The time $t$ is
expressed in units of $\frac{\Omega }{2\pi }.$

We have also calculated numerically and plotted as a function of $\alpha
_{r} $ and $\alpha _{i}$ the Quasi-Probability function $Q\left( \alpha
\right) $ defined as 
\begin{equation}
Q(\alpha )=\frac{1}{\pi }_{q}\left\langle \alpha \right| \rho
_{c.m.}(t)\left| \alpha \right\rangle _{q}\text{ ,}  \label{eq26}
\end{equation}
where $\rho _{c.m.}(t)=tr_{ion}\rho (t)$ is the density matrix reduced for
degrees of freedom of center of mass motion.

\section{Quantum Mutual Entropy}

Dynamical behavior of the system can be better understood in terms of
quantum mutual entropy. Furuichi and co-workers\cite{fur99} have applied the
concept of quantum mutual entropy to dynamical change of state of the atom
in Jaynes-Cummings model(JCM)\cite{Jayn63}. The analogy between JCM and ion
in a trap system allows us to extend the concept to understand the
entanglement of the internal degrees of freedom of ion and it's motional
degrees of freedom. We next outline the calculation of quantum mutual
entropy for the system at hand prepared initially in state given by Eq. (\ref
{eq20}). It is found that the quantum mutual entropy can be decomposed into
a part determined by the relative populations of the system and another that
depends on the coherences. The partial quantum mutual entropy used by us in
our numerical study is essentially a measure of mutual entropy due to
populations developed in the system with the passage of time.

With the initial state of the trapped ion expressed as in (\ref{eq20}), The
density operator for the internal states of the ion at $t=0$ is 
\begin{equation}
\rho _{ion}=\lambda _{1}\left| g\right\rangle \left\langle g\right| +\lambda
_{2}\left| e\right\rangle \left\langle e\right| .  \label{eq27}
\end{equation}
The state of motion of the ionic center of mass at $t=0$ is represented by a
coherent state $\left| \beta \right\rangle _{q}$ for ion in ground state and
by coherent state $\left| -\beta \right\rangle _{q}$ for ion in excited
state. where 
\begin{equation}
\left| \beta \right\rangle _{q}=\frac{1}{\sqrt{\exp _{q}\left( \left| \beta
\right| ^{2}\right) }}\sum\limits_{n=0}^{\infty }\frac{\beta ^{n}}{\sqrt{%
\left[ n\right] _{q}!}}\left| n\right\rangle _{q}\quad .  \label{eq28}
\end{equation}
The state of the system at a time $t$ is given by Eq. (\ref{eq22}).

The time evolution of the internal state of the ion can also be represented
by a continuous mapping 
\begin{equation}
\Lambda _{t}^{ion}\rho _{ion}=tr_{c.m.}\left( U_{t}\rho _{0}U_{t}\right)
\label{eq29}
\end{equation}
where $\rho _{0}$ is the initial state of the system, $U_{t}=\exp \left( 
\frac{-iHt}{\hbar }\right) $ a unitary operator and the mapping $\Lambda
_{t}^{ion}$ maps the initial internal state $\rho _{ion}$ to the current
internal state of the ion at time $t$. We represent the time evolution of
initial constituent ionic states $\left| g\right\rangle \left\langle
g\right| $ and $\left| e\right\rangle \left\langle e\right| $ through the
maps 
\begin{eqnarray}
\Lambda _{t}^{ion}\left| g\right\rangle \left\langle g\right|
&=&tr_{c.m.}\left( U_{t}\left| g,\beta \right\rangle \left\langle g,\beta
\right| U_{t}\right) =P_{g}^{(1)}(t)\left| g\right\rangle \left\langle
g\right| +P_{e}^{(1)}(t)\left| e\right\rangle \left\langle e\right|
\label{eq30} \\
&&+C_{ge}^{(1)}(t)\left| g\right\rangle \left\langle e\right|
+C_{ge}^{(1)^{*}}(t)\left| e\right\rangle \left\langle g\right|
\end{eqnarray}
and 
\begin{equation}
\Lambda _{t}^{ion}\left| e\right\rangle \left\langle e\right|
=tr_{c.m.}\left( U_{t}\left| e,-\beta \right\rangle \left\langle e,-\beta
\right| U_{t}\right) \text{ .}  \label{eq31}
\end{equation}
The correlation between the initial ionic state and its current state
containing information about the time evolution of each constituent ionic
states is given in terms of the quantum mutual entropy defined as 
\begin{equation}
S_{m}(\rho ^{ion},\Lambda _{t}^{ion})=\sum_{i=1}^{2}\lambda _{i}S\left(
\Lambda _{t}^{ion}\left| i\right\rangle \left\langle i\right| ,\quad \Lambda
_{t}^{ion}\rho _{ion}\right) \text{ ,}  \label{eq32}
\end{equation}
where the relative quantum entropy 
\begin{eqnarray}
S\left( \Lambda _{t}^{ion}\left| i\right\rangle \left\langle i\right| ,\quad
\Lambda _{t}^{ion}\rho _{ion}\right) &=&tr\left( \Lambda _{t}^{ion}\left|
i\right\rangle \left\langle i\right| \right) \left[ \log \left( \Lambda
_{t}^{ion}\left| i\right\rangle \left\langle i\right| \right) \right. 
\nonumber \\
&&\left. -\log \left( \Lambda _{t}^{ion}\rho _{ion}\right) \right]
\label{eq33}
\end{eqnarray}
and $i=1$ $(2)$ for ion in ground (excited) state. It measures the relative
overlap of the constituent ionic state $\left| i\right\rangle $ with the
ionic state of the system at a time $t$ for the cases a) where the initial
ionic state of the system is state $\left| i\right\rangle $ and b) the
initial ionic state of the system is given by Eq. (\ref{eq27}). The quantum
relative entropy is a more general concept than the Von Neumann Entropy
being a measure that extends to mixed states. As defined in Eq. (\ref{eq33}%
), it is a measure of how far each entangled state is from its component
disentangled state as a result of time evolution. The quantum mutual entropy
as defined by Eq. \ref{eq32} is obtained by summing up the time dependent
quantum relative entropies with respect to ionic states that constitute the
initial cat state and measures the propagation of information content of the
initial state.

Working in the space spanned by basis vectors $\{\left| u_{n}\right\rangle
\}\equiv \left| g,m\right\rangle ,\left| e,m\right\rangle ,m=0,1,.....\infty
,$ we find from Eq. (\ref{eq22}) the probabilities of finding the ion in the
ground state and the excited state (the populations) to be respectively 
\begin{equation}
P_{g}(t)=\left\langle g\left| tr_{ion}\rho (t)\right| g\right\rangle
=\sum_{n}\left| g_{n}(t)\right| ^{2}  \label{eq34}
\end{equation}
and 
\begin{equation}
P_{e}(t)=\left\langle e\left| tr_{ion}\rho (t)\right| e\right\rangle
=\sum_{n}\left| e_{n}(t)\right| ^{2}.  \label{eq35}
\end{equation}
Besides that the system at a time $t$ is characterized by the following
non-diagonal matrix element of the operator $tr_{ion}\rho (t)$ (coherence), 
\begin{equation}
C_{ge}=\left\langle g\left| tr_{ion}\rho (t)\right| e\right\rangle
=\sum_{n}g_{n}^{*}(t)e_{n}(t)\qquad .  \label{eq36}
\end{equation}
We also define the populations and coherences for the system prepared in
initial state $\left| g,\beta \right\rangle _{q}$ ($i=1$) and the initial
state with the ion in the excited state $\left| e,-\beta \right\rangle _{q}$
($i=2$). The state of the system at a time $t$ is now given by 
\begin{equation}
\Psi ^{(i)}(t)=\sum\limits_{m}g_{m}^{(i)}(t)\left| g,m\right\rangle
+\sum\limits_{m}e_{m}^{(i)}(t)\left| e,m\right\rangle \quad ,  \label{eq37}
\end{equation}
with the populations defined as 
\begin{equation}
P_{g}^{(i)}(t)=\left\langle g\left| tr_{ion}\rho _{{}}^{(i)}(t)\right|
g\right\rangle =\sum_{n}\left| g_{n}^{(i)}(t)\right| ^{2}  \label{eq38}
\end{equation}
and 
\begin{equation}
P_{e}^{(i)}(t)=\left\langle e\left| tr_{ion}\rho _{{}}^{(i)}(t)\right|
e\right\rangle =\sum_{n}\left| e_{n}^{(i)}(t)\right| ^{2}.  \label{eq39}
\end{equation}
The coherences for these initial states are given by 
\begin{equation}
C_{ge}^{(i)}(t)=\left\langle g\left| tr_{ion}\rho _{{}}^{(i)}(t)\right|
e\right\rangle =\sum_{n}\left( g_{n}^{(i)}(t)\right)
^{*}e_{n}^{(i)}(t)\qquad .  \label{eq40}
\end{equation}
We can verify that in terms of the populations and the coherences defined
above the relative quantum entropy, $S\left( \Lambda _{t}^{ion}\left|
i\right\rangle \left\langle i\right| ,\quad \Lambda _{t}^{ion}\rho
_{ion}\right) $ (Eq. \ref{eq33}) for the system starting in initial state
given by Eq. (\ref{eq27}) is 
\begin{eqnarray}
S\left( \Lambda _{t}^{ion}\left| i\right\rangle \left\langle i\right| ,\quad
\Lambda _{t}^{ion}\rho _{ion}\right) &=&\left[ P_{g}^{(i)}(t)\log \left( 
\frac{P_{g}^{(i)}(t)}{P_{g}^{{}}(t)}\right) +P_{e}^{(i)}\log \left( \frac{%
P_{e}^{(i)}(t)}{P_{e}^{{}}(t)}\right) \right.  \nonumber \\
&&\left. +2*%
\mathop{\rm Re}%
\left[ C_{ge}^{(i)}(t)\log \left( \frac{C_{ge}^{(i)}(t)}{C_{ge}^{{}}(t)}%
\right) ^{*}\right] \right] \text{ .}  \label{eq41}
\end{eqnarray}
Substituting Eq. (\ref{eq41}) in the definition of quantum mutual entropy
for the system, we have 
\begin{eqnarray}
S_{m} &=&\sum_{i=1,2}\lambda (i)\left[ P_{g}^{(i)}(t)\log \left( \frac{%
P_{g}^{(i)}(t)}{P_{g}^{{}}(t)}\right) +P_{e}^{(i)}\log \left( \frac{%
P_{e}^{(i)}(t)}{P_{e}^{{}}(t)}\right) \right.  \nonumber \\
&&\left. +2*%
\mathop{\rm Re}%
\left[ C_{ge}^{(i)}(t)\log \left( \frac{C_{ge}^{(i)}(t)}{C_{ge}^{{}}(t)}%
\right) ^{*}\right] \right] \qquad .  \label{eq42}
\end{eqnarray}
We can split $S_{m}$ in to two distinct parts, one depending on populations
and the other on off diagonal coherences that is 
\begin{equation}
S_{m}=S(P)+S(C)  \label{eq43}
\end{equation}
where 
\begin{equation}
S(P)=\sum_{i=1,2}\lambda (i)\left[ P_{g}^{(i)}(t)\log \left( \frac{%
P_{g}^{(i)}(t)}{P_{g}^{{}}(t)}\right) +P_{e}^{(i)}(t)\log \left( \frac{%
P_{e}^{(i)}(t)}{P_{e}^{{}}(t)}\right) \right] \text{ .}  \label{eq44}
\end{equation}
We have calculated numerically $S(P)$ the partial mutual quantum entropy for
the cases $\beta =3,4$ and used the $S(P)$\ peaks to pinpoint the $t$ values
for which the system is the most correlated and the $t$ values where it
shows the least correlation. Essentially $S(P)$ is used as a measure of
entanglement of the system.

\section{Results}

The population inversion as a function of rescaled time parameter $t(\frac{%
\Omega t}{2\pi })$ for two level ion interacting with light field in a
harmonic oscillator trap and a $q$-deormed oscillator trap for the cases $%
\beta =3,4$ is displayed in Figures. 1 and 3 for three different values of
deformation parameter $\tau $ ($q=\exp (\tau )$ with $\tau $ $%
\mathop{\rm real}%
$). The set of parameters used in the numerical calculation is $\overline{%
\omega }=\frac{\omega }{\Omega }=50,$ $\overline{\epsilon }=\frac{\epsilon }{%
\Omega }=0.05,$ and $\overline{\Delta }=\frac{\Delta }{\Omega }=-50$. The
maximum value of $n$ in Eq. (\ref{eq22}) is restricted to $n=32$. For both $%
\beta $ values the collapse and revival pattern was found to improve in
definition with increase in the anharmonicity of the trap potential,
measured by parameter $\tau $. In Figures. 1a and 2a, we display the
collapse and revival for ion in a harmonic oscillator trap that is $\tau
=0.0.$ Both for $\beta =3$ and $4$ first revival and second revival are
easily identified. For $\beta =3$ the second collapse, which is barely
identifiable is succeeded by a region where the inversion pattern becomes
diffuse, whereas for $\beta =4$ the second collapse is followed by another
revival which is quite wide. For $\beta =3$ and $4,$ the best collapse and
revival sequence is obtained for $\tau =0.0047$ and $\tau =0.004$
respectively as shown in Figures. 1b and 2b. We can identify four revivals
in each case. The collapses and revivals for $\beta =4$ are remarkably
sharp. Further increase in $\tau $ results in a loss of contrast between
successive revival peaks. Figures. 1c and 2c indicate a rapid loss of phase
relationship essential for generating observable collapses and revivals of
population inversion between components of the total wave function as the
anharmonicity of the trap potential is increased. We have also found that
for a large value of $\tau ,$ different for each $\beta $ value, not only
the collapses and revivals of population inversion vanish but also the time
dependence of population inversion disappears in all cases.

In figs. (3-4) we plot the partial quantum mutual entropy $S(P)$ as a
function of time for initial states with $\beta =3,$ and $4.$ A careful
examination shows that each collapse as well as revival of population
inversion is characterized by a peak in $S(P)$ versus $t$ plot. During the
transition from collapse to revival and vice-versa we have minimum mutual
entropy value that is $S(P)=0.$ In the $q$-deormed oscillator trap for the
cases $\beta =3,\tau =0.0047$ and $\beta =4,\tau =0.004$ we notice
relatively higher $S(P)$ peaks and well defined transition periods(collapse $%
\Leftrightarrow $ revival) in comparison with the harmonic oscillator trap($%
\tau =0.0)$. The regions without continuous minimum $S(P)$ do not show well
defined collapses or revivals in the population inversion plots. In Table I,
the $t$ values for successive peaks in $S(P)$ for the cases $\beta =3,\tau
=0.0047$ and $\beta =4,\tau =0.004$ are displayed. It is evident from
figures 3 and 4 that $S(P)$ decreases with the Rabi oscillations. Table I
demonstrates that the successive revival peaks show a lowering of the local
maximum point indicating a dissipative irreversible change in the ionic
state.

\begin{tabular}{l}
Table I. The values of $t$ for successive peaks in $S(P)$ for the cases \\ 
$\beta =3,\tau =0.0047$, $\beta =4,\tau =0.0$ and $\beta =4,\tau =0.004.$%
\end{tabular}

\begin{tabular}{|r|r|r|r|r|r|r|r|r|}
\hline
\multicolumn{3}{|r|}{$\beta =3,\tau =0.0047$} & \multicolumn{3}{|r}{$\beta
=4,\tau =0.0$} & \multicolumn{3}{|r|}{$\beta =4,\tau =0.004$} \\ \hline
$t$ & $S(P)$ & Peak type & $t$ & $S(P)$ & Peak type & $t$ & $S(P)$ & Peak
type \\ \hline
$0.0$ & $0.693$ & Initial state & $0.0$ & $0.693$ & Initial state & $0.0$ & $%
0.693$ & Initial state \\ \hline
$58.6$ & $0.628$ & revival & $85.8$ & $0.333$ & revival & $67.8$ & $0.731$ & 
revival \\ \hline
$114.0$ & $0.314$ & collapse & $171.4$ & $0.143$ & collapse & $133.2$ & $%
0.497$ & collapse \\ \hline
$175.6$ & $0.285$ & revival & $266.8$ & $0.106$ & revival & $201.2$ & $0.512$
& revival \\ \hline
$234.0$ & $0.220$ & collapse & $388.2$ & $0.085$ &  & $266.6$ & $0.359$ & 
collapse \\ \hline
$295.6$ & $0.181$ & revival & $447.6$ & $0.076$ &  & $336.8$ & $0.397$ & 
revival \\ \hline
$360.0$ & $0.234$ & collapse &  &  &  & $404.8$ & $0.334$ & collapse \\ 
\hline
$415.4$ & $0.175$ & revival &  &  &  & $472.6$ & $0.324$ & revival \\ \hline
\end{tabular}

Next we plot in Fig. 5 the Quasi-probabilities for $\beta =4$ at $%
t=0.0,85.8,171.4,266.8,388.2$ and $447.6$ to visualize the time evolution of
ionic center of mass in a harmonic oscillator trap($\tau =0.0$). As
expected, the coherent states $\left| g,\beta \right\rangle _{{}}$ and $\
\left| e,-\beta \right\rangle $ , each splits into two and move in a sense
opposite to each other in phase space. At $t=85.8$ there are two somewhat
distorted coherent states characterized by almost equal average position
coordinate($\alpha _{r}\simeq 0)$, and momenta with opposite signs. This
point corresponds to a revival peak in the population inversion plot and the
first peak S(P) versus $t$ plot( Fig. 4a). A second revival similarly occurs
at $t=266.8$. The collapse at $t=171.4$ corresponds to a situation similar
to the initial state. At $388.2$ and $447.6$ the quasi-probability is spread
out in the phase space. For $\tau =0.004$ quasi-probability contour plots at 
$t=0.0,67.8,133.2,201.2,266.6,$ and $336.8$ are presented in Fig. 6.
Characteristic spreading of the Quasi-probabilities in the phase space is
the result of non-linearities of the trap, however the initial state
configuration is still recovered signalling collapse and revival phenomenon.

\section{Conclusion}

Entangled quantum systems present correlations having no classical
counterparts and may be exploited to store and transfer information in ways
otherwise difficult or impossible. In this work we have examined the effects
of trap non-linearity on Schr\"{o}dinger cat state of a single trapped ion.
This particular quantum system may serve in future as a basic element in
quantum computation and quantum communication \cite
{Benn92,Benn93,Eker91,Eker92} or a sensitive detector of decoherence due to
interaction with environment. The entanglement of ionic internal states with
center of mass motion may be used to encode information in such
applications. We have found an enhancement in the contrast between the
collapse and revival peaks as the anharmonicity of the trapping potential
increases. With further increase in non-linearity not only the collapses and
revivals disappear but also the time dependence of population inversion
vanishes. This extremely interesting feature is manifested in the Population
inversion plots, the Quasi- probability plots as well as the partial mutual
entropy plots. The effect is markedly visible in $\beta =4$ case for the set
of parameters used to characterize the ion-laser in trap system. The best
collapse and revival sequences are obtained for $\tau =0.0047$ and $\tau
=0.004,$ for $\beta =3$ and $4$ respectively. The reason for this effect is
to be found in the deformation dependence of trap energies. The
characteristic Rabi Frequency, in addition to the ion-laser interaction now
depends on energy separation between trap levels. It can be easily verified
that in the lowest approximation, with $\omega \gg \Omega $ and driving
field tuned to $\Delta =-\omega ,$the Rabi frequency is given by $\mu (n)=%
\sqrt{\left( \frac{\omega }{2}(\cosh (2(n+1)\tau )-1)\right) ^{2}+\Omega
^{2}\epsilon ^{2}[n+1]_{q}}$ which in the limit $\tau \rightarrow 0$ (for $%
q=e^{\tau }$) reduces to $\mu (n)=\sqrt{\Omega ^{2}\epsilon ^{2}[n+1]}$ as
expected. For small values of parameter $\tau ,$ the factor $\left( \frac{%
\omega }{2}(\cosh (2(n+1)\tau )-1)\right) ^{2}$, coming from non-equidistant
trap energy spectrum, varies approximately as $(n+1)^{4}$ and dominates the
scene resulting in a high contrast collapse and revival peaks in population
inversion. For large values of $\tau $ decoherence sets in accompanied by
loss of amplitude of population inversion and for $\tau $ $\sim 0.1$ the
collapse and revival phenomenon disappear. We use partial quantum mutual
entropy as an indicator of time evolution of entanglement. It is obtained by
summing up the time dependent quantum relative entropies with respect to
ionic states that constitute the initial cat state and retaining the part
that depends on populations. Quantum relative entropy is a measure of how
far each entangled state moves from its component disentangled state as a
result of time evolution. Looking together at the population inversion and
partial mutual entropy $S(P)$ plots we verify that the onset of collapse and
revival is characterized by $S(P)=0$. A well defined collapse and revival
pattern is characterized by wide $S(P)=0$ regions. Every collapse or revival
corresponds to a peaked $S(P)$. Successive revival peaks show a lowering of
the local maximum point which is an indicator of a dissipative irreversible
change in the ionic state same being true for $S(P)$ peaks that indicate
collapse of the population inversion.

{\Large Acknowledgments}

S. S. S. and N. K. S. would like to acknowledge financial support from Funda%
\c{c}\~{a}o Araucaria, PR, Brazil and CNPq, Brazil.

\input epsf \newpage \epsfysize=4.9in \center{\epsfbox{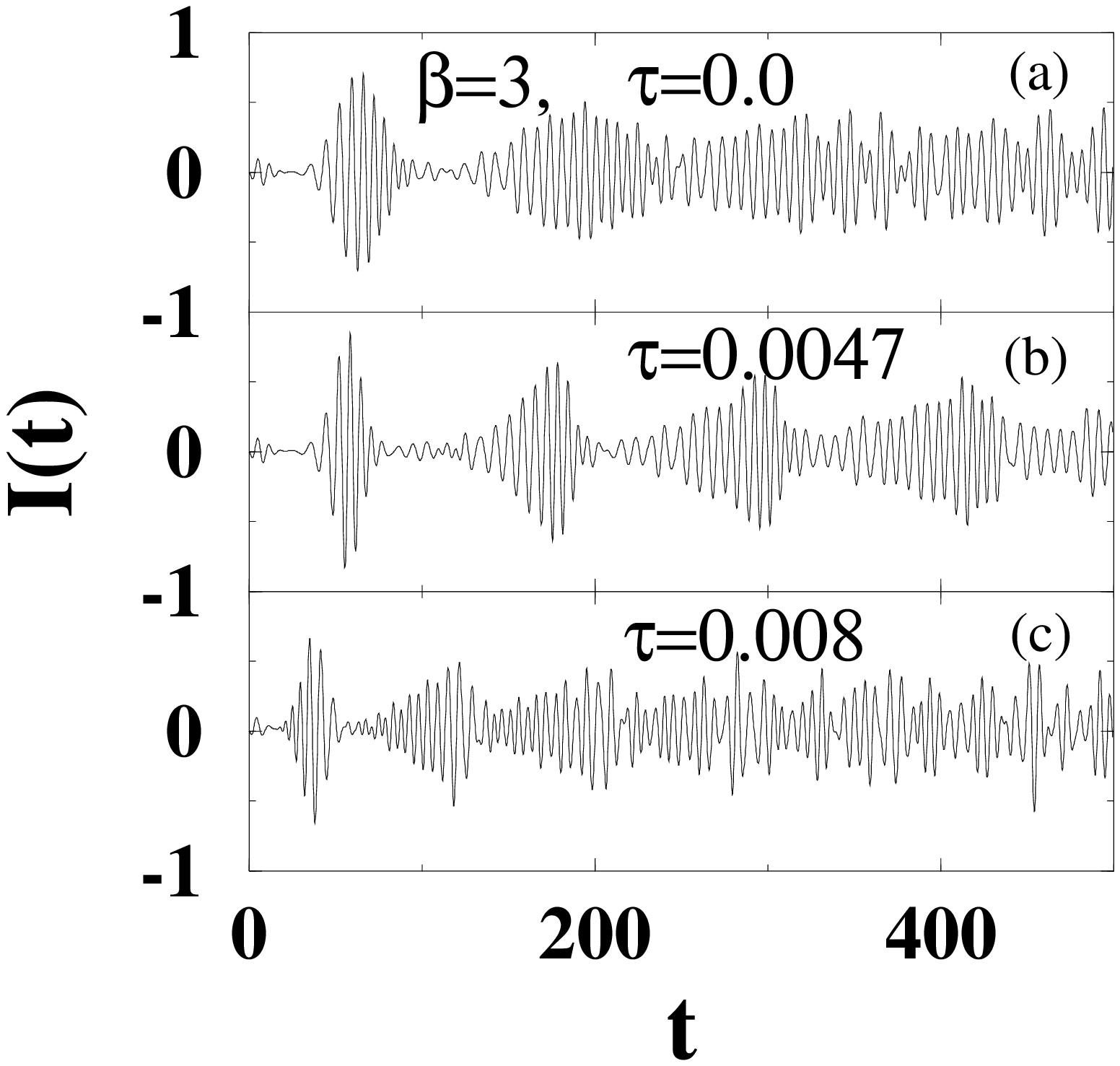}}

\center{Fig. 1. {Population Inversion $I(t)$ versus $t$ for $\beta =3$ and $%
\tau=0.0$, $0.0047$, $0.008$.}}

\newpage \epsfysize=4.9in \epsfbox{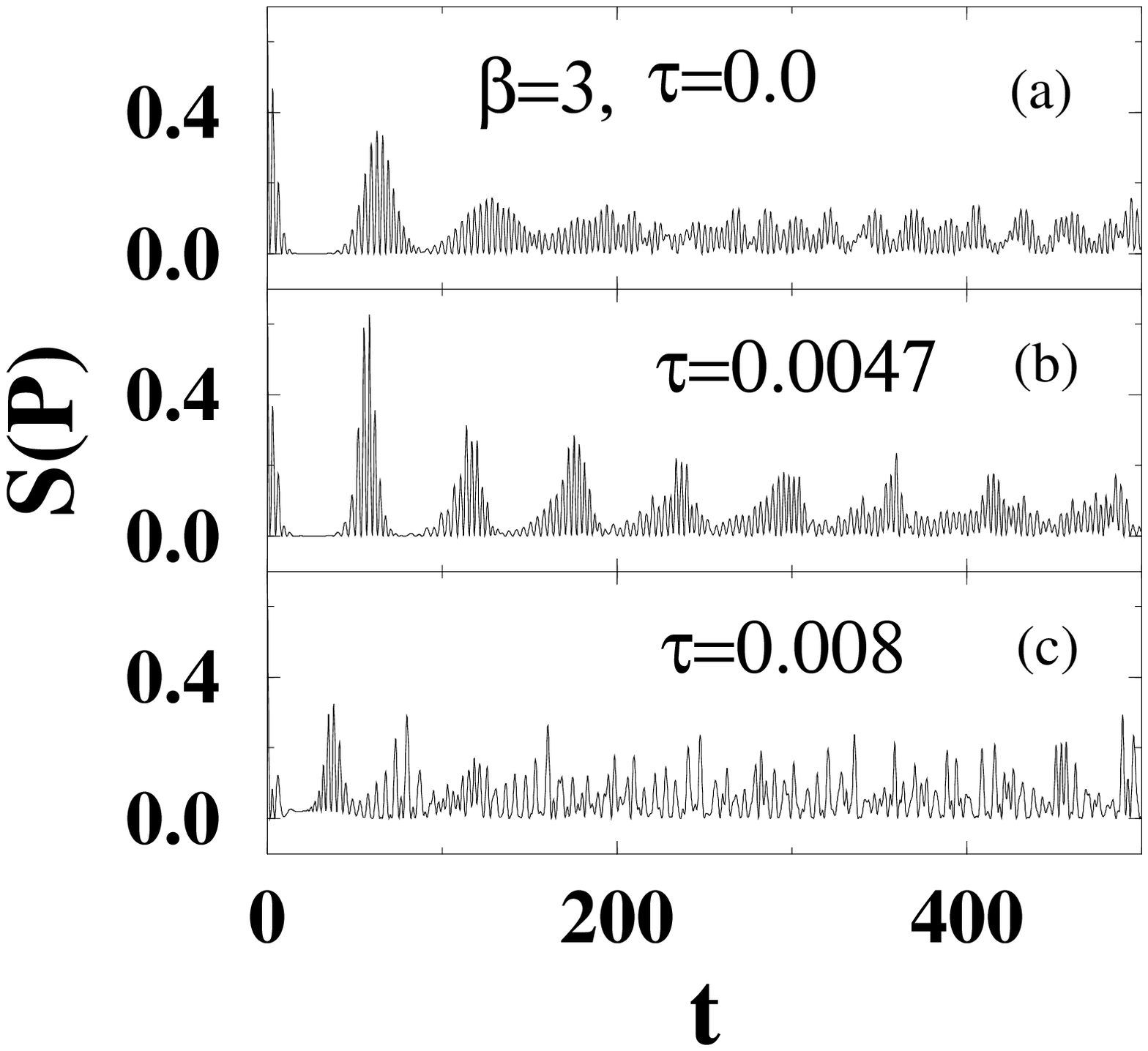}

\center{Fig. 2. {Partial mutual quantum entropy $S(P)$ versus $t$ for$\beta
=3$ and $\tau =0.0, 0.0047, 0.008$.}}

\newpage \epsfysize=4.9in \epsfbox{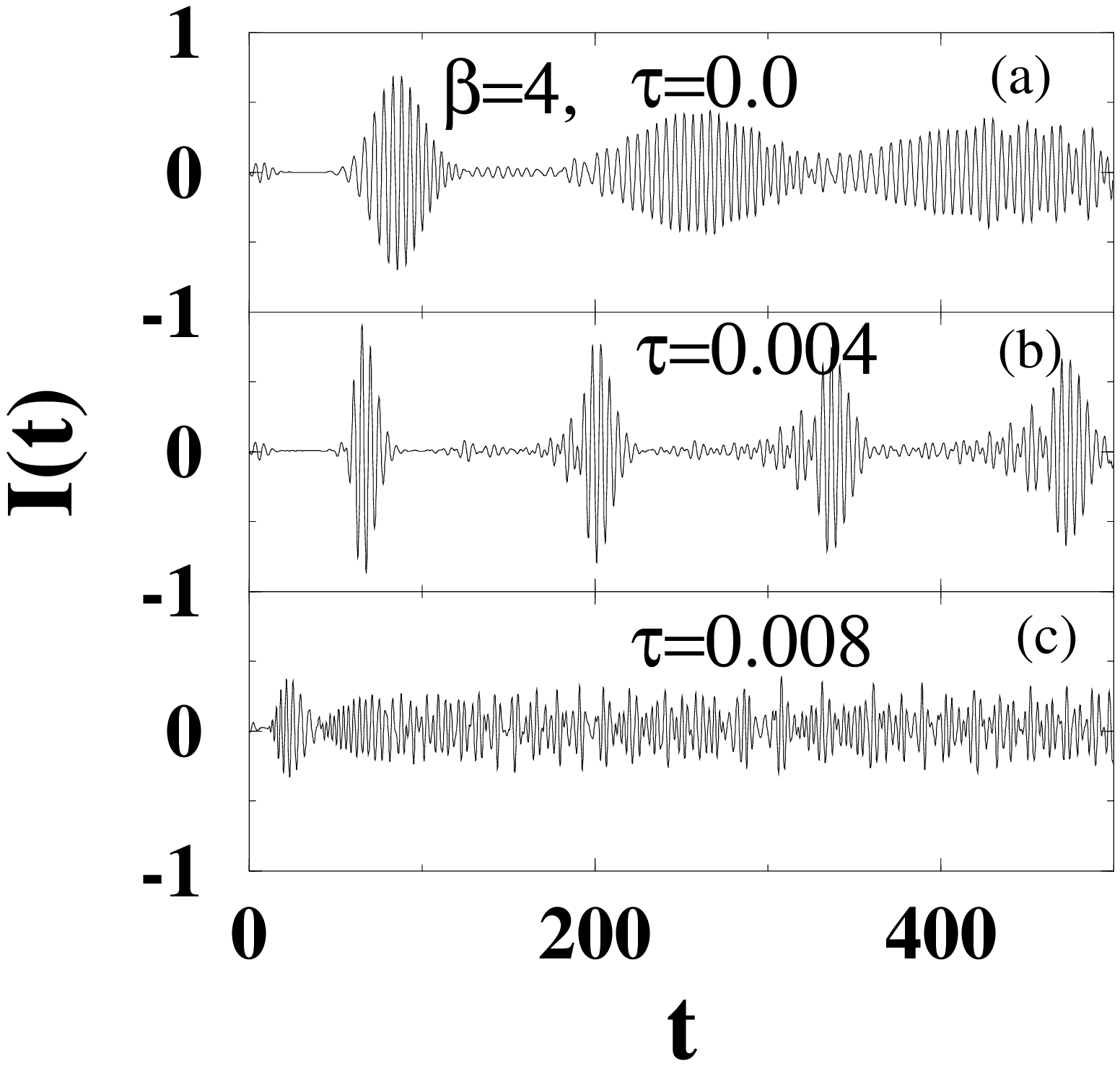}

\center{Fig. 3. {Population Inversion $I(t)$ versus $t$ for $\beta =4$ and $%
\tau=0.0, 0.004, 0.008$.}}

\newpage \epsfysize=4.9in \epsfbox{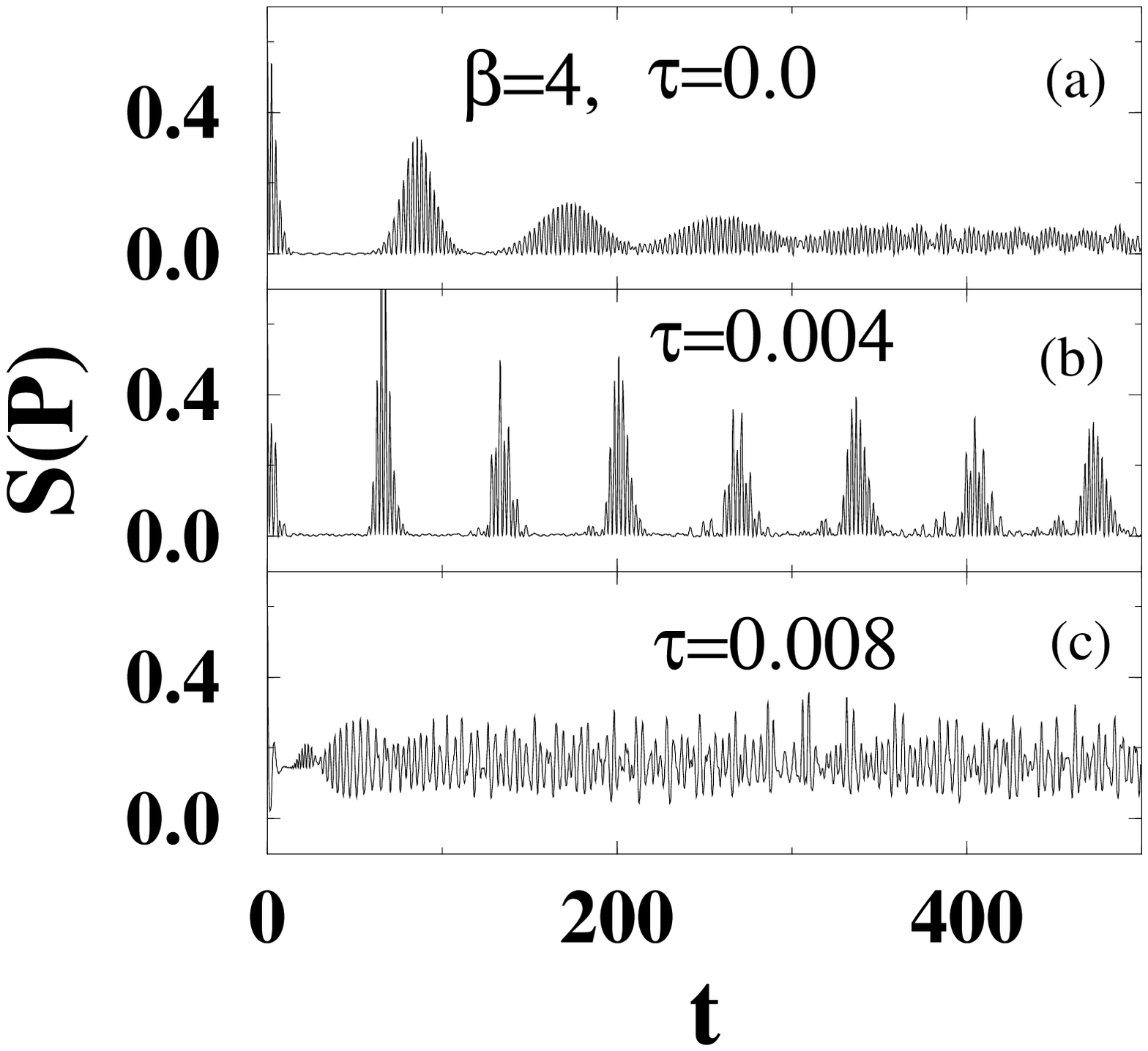}

\center{Fig. 4. {Partial mutual quantum entropy $S(P)$ versus $t$ for $\beta
=4$ and $\tau =0.0, 0.004, 0.008$.}}

\newpage \epsfxsize=5.4in \epsfbox{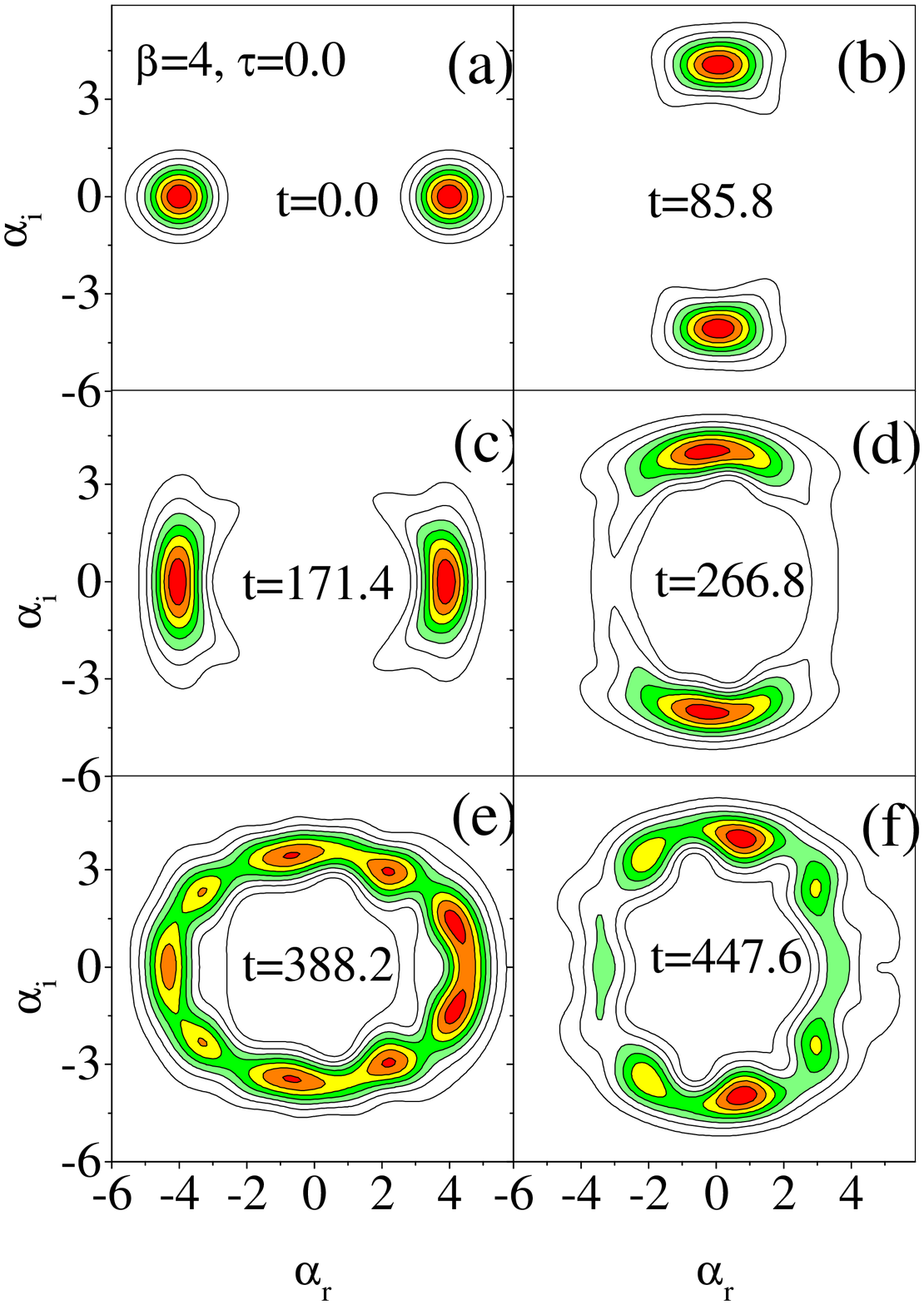}

Fig. 5. {Quasi-probability plots for $\beta =4$ and $\tau =0.0$ at$t=0.0$,$%
85.8$,$171.4$,$266.8$,$388.2$ and $447.6$.}

\newpage \epsfxsize=5.4in \epsfbox{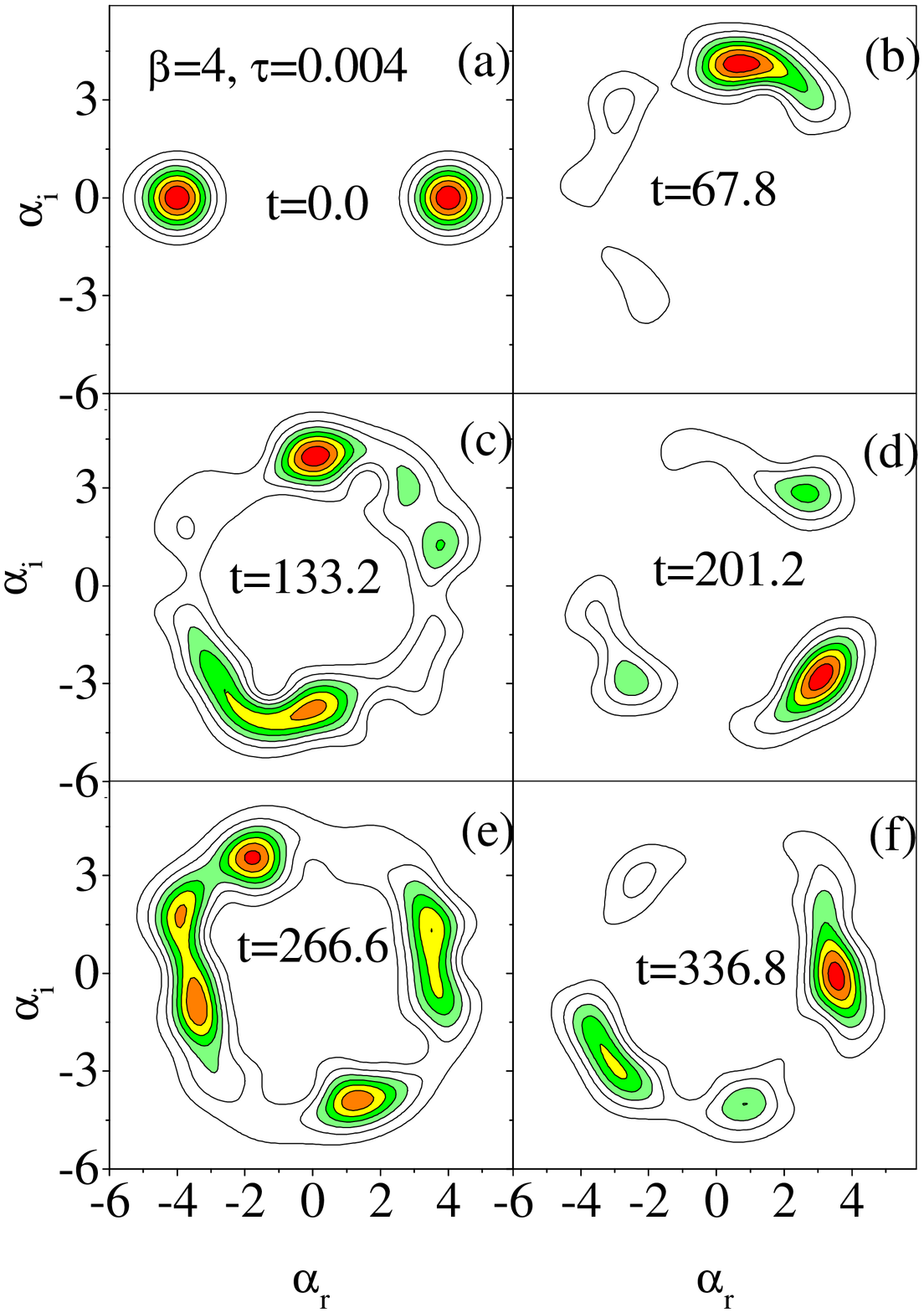}

Fig. 6. {Quasi- probability plots for $\beta =4$ and $\tau =0.004$ at$t=0.0,
67.8, 133.2, 201.2, 266.6,$ and $336.8$.}

\end{document}